\documentclass[12pt,thmsa]{article}

\begin{document}

\author{Dario Sassi Thober \and Center for Research and Technology CPTec - Unisal \and thober@fnal.gov thober@cptec.br \and Phone.55-19-7443107 Fax.55-19-7443036}
\title{On the Dirac monopole's concept}
\date{}
\maketitle

\begin{abstract}
The Dirac monopole is discussed in view of the gauge invariance in Quantum
Electrodynamics. It is shown the monopole existence implies the violation of
the gauge invariance principle. The monopole field is essentially a
longitudinal field and so a mass is naturally associated to it. Interpretation for the case the monopole  charge is different from zero is addressed at the conclusion. 
\end{abstract}

\section{The topologycal monopole}

Any derivation of the Dirac Quantization condition [1], $eg/c=n\hbar /2$
(where $e$ is the electric charge, $g$ the magnetic flux and $n=0,\pm 1,\pm
2,...$), is based on two steps: First, one has to calculate the angular
momentum due to the fields; Second, to quantize it through Schroedinger
Quantum Mechanics (for instance).

The first step was first performed in 1904 when J. J. Thomson [2] derived
the term $e\mathbf{v\times B}$ known as the magnetic part of the Lorentz
force upon an electric charge ($\mathbf{B}$ is the magnetic field and $%
\mathbf{v}$ the charge's velocity, both relative to the inertial laboratory
system). The angular momentum of the field resulting from an electric charge
and a magnetic monopole (or simply, monopole) can be written as the volume
integral of the angular momentum density $\mathbf{r\times (E\times B)}/4\pi
c $, with $\mathbf{r}=(x,y,z)$ (the position vector) and $\mathbf{E}$ the
electric field. It results the total angular momentum stored in the field
for this system is $eg/c$. Considering the angular momentum part due to the
particles has to compensate the one of the fields, Thomson derived the
magnetic Lorentz term which leads to a variation of the angular momentum of $%
2eg/c$ for the particles in order to preserve the conservation of the total
angular momentum of the system.

The second step towards the definition of the Dirac Quantization Condition
is to consider Wave Mechanichs as valid for particles, [1]. Any gauge change
on the electromagnetic potentials leaves the Schroedinger equation invariant
if the wave functions are altered as $\psi \rightarrow \psi ^{\prime }=\psi
\exp \left[ ie\phi /\hbar c\right] $, where $\phi $ is the gauge function.
It is clear that, once the phase $e\phi /\hbar c$ has no dimension, $\phi $
is directly connected to some magnetic flux. In the system of one charge and
one monopole this magnetic flux is $4\pi g$, and requiring the phase must be
unaltered by the presence of the monopole, i.e., that the phase change due
to monopole's presence must be a multiple of $2\pi $, one gets the Dirac
Quantization Condition, [1]. It has been considered this condition makes the
gauge due to the monopole just an artifact. Accordingly, it was shown by
other approaches (Wu and Yang's, [3]) that if such condition is observed the
monopole is a topologycal set-up of the gauge fields, being the monopole's
gauge not observable.

It is interesting to quote this condition was derived by Dirac [1] and
Wu-Yang [3] in order to the charges never cross a singularity on the vector
potential gauge associated to the monopoles. In Brandt and Primack's work,
[4], it was shown also that the two approaches are equivalent. In Dirac's
approach the wave functions of the charges are set to be exactly null on the
singular part of the vector field, while in Wu-Yang's a Fiber-Bundle
description is given as the traduction of this condition. In fact, it is
impossible to define a monopole through a vector potential (from which the
magnetic field is to be derived) free of singularities in $\mathbf{R}^{3}$
(the three-dimensional flat world), [3], and Dirac Quantization Condition
makes the singularity non-observable.

As the approaches of Dirac, [1] and Wu-Yang, [3], are equivalent, [4], we
proceed considering the consequences in assuming the existence of a monopole 
$g$ and the condition $eg/c=n\hbar /2$. He, Qiu and Tze, [5], considered
this situation and showed it is impossible to get $g\neq 0$ in pure Quantum
Electrodynamics (QED). It can be demonstrated that the longitudinal photon
field does not alter Physics in a gauge-invariant theory as QED, [5], so the
unphysical gauge coupling (which has an arbitrary value) associated to this
field is not observable. It was shown however, [5], that in deriving the
Dirac Quantization Condition, this unphysical gauge coupling appears related
to the monopole flux in the same way it appears for the physical electric
charge in the usual relation $eg/c=n\hbar /2$. This is due to the fact the
non-integrable phase for the particle with charge $e$ in the presence of the
fields $A_{\mu }$ (physical) and $[A_{0}]_{\mu }$ (unphysical logitudinal
field) can be written as: 
\begin{equation}
P(x_{2},x_{1};C)=\exp \left[ i\frac{e}{\hbar c}\int_{x_{1}}^{x_{2}}A_{\mu
}(x)dx^{\mu }+i\frac{e_{0}}{\hbar c}\int_{x_{1}}^{x_{2}}[A_{0}]_{\mu
}(x)dx^{\mu }\right] 
\end{equation}
where the line-integral is along the path $C$ from $x_{1}$ to $x_{2}$. For a
closed loop $C$, equation (1) gives: 
\begin{equation}
P(x,x;C)=\exp \left[ i\frac{e}{\hbar c}\int_{a}F_{\mu \nu }da^{\mu \nu
}\right] 
\end{equation}
where $F_{\mu \nu }$ is the electromagnetic field strenght and $a$ is the
area bounded by $C$. Now $P$ is $U(1)$ gauge-invariant as the unphysical
field $[A_{0}]_{\mu }$makes no contribution, [5].

Now consider we are going to measure the magnetic flux over the area $a$, $%
\int \mathbf{B\cdot da}$, knowing that $\mathbf{B=\nabla \times A}$, and
that $\mathbf{A=A+\nabla }\varphi $, where $\varphi $ represents the
gauge-freedom for the vector potential. The measure of the magnetic flux can
be written as: 
\begin{equation}
\int \mathbf{\nabla \times A\cdot da}
\end{equation}

According to Wu and Yang, [3], for each point over the surface $a$ we must
choose $\mathbf{A}$ such that no singularity is visited on going over the
surface, i.e., if some point which will be considered is the place occupied
by the singular string on this surface, the gauge changing (from one
previous point to this) must respect the condition $eg/c=n\hbar /2$ (a $2\pi 
$ change of phase), in a way that the string is moved to another place in $%
\mathbf{R}^{3}$. This way no singularity is seen, the term associated to the
longitudinal field makes no contribution and the Dirac Quantization
Condition is respected, [3].

According to reference [5], this is wrong. It is clear that the
tranformation from equation (1) to (2) is only possible if the Stokes
theorem can be applied, and unfortunately, $\mathbf{A}$ is not a continuous
function to support such theorem. It changes abruptly from one point to the
next in space due to the gauge changing necessary to hide the singularity
associated to the monopole's definition. The Stokes theorem cannot be
applied, therefore the unphysical gauge coupling $e_{0}$ cannot be
disregarded and will appear connect to the magnetic flux $4\pi g$ as $%
e_{0}g/c=n\hbar /2$, [5]. It is easy to see that if primary the gauge is $%
\mathbf{A}_{1}=\mathbf{\phi }g(1-\cos \theta )/r\sin \theta $, for $\theta
<\pi -\varepsilon $, and changes to $\mathbf{A}_{2}=-\mathbf{\phi }g(1+\cos
\theta )/r\sin \theta $, for $\theta >\varepsilon $, for some electric
charge's path in the region of the overlap of the two gauges ($\varepsilon $
is some arbitrary angle), one goes suddenly from $G(1-Y)\mathbf{\phi }$ to $%
G(-1-Y)\mathbf{\phi }$ (if $G\equiv g/r\sin \theta $, and $Y\equiv \cos
\theta $). The only possible solution for this situation is to set $g=0$ as
in reference [5], otherwise a physical interpretation must be assumed for $%
e_{0}$ in pure QED since we get both $eg/c=n\hbar /2$ and $e_{0}g/c=n\hbar
/2 $.

In the next section an alternative approach to such problem is addressed. It
is shown the gauge invariance is broken because longitudinal fields are
related to the monopole's definition. We then proceed considering the case $
g\neq 0$ and an interpretation is addressed. If the string of sigularities
is not to be seen, some physical reason (strange to pure QED) must be given
for this to happen.

\section{The definition of a Dirac monopole}

In the following steps we will start from the Maxwell equations in the
vacuum and consider the necessary conditions for the existence of a monopole
with $g\neq 0$. We do so because the definition $\mathbf{B}=\mathbf{\nabla }%
\times \mathbf{A}$ was first derived from the condition $\mathbf{\nabla
\cdot B}=0$. The departure equations are those which describe photons in the
usual theory (gauge invariant). At the end, after defining the composition
of fields and conditions which define the monopole, one desires part of the
flux of its field (the returning flux) to be associated to the arbitrary
gauge of the system in a way the definition of the monopole is free of
singularities (the returning flux) in a topologycal manner, [3]. In the next
steps it will become clear that in order to define a monopole, the gauge
invariance must be broken and as a consequence this violation is not only
connected to the magnetic charge but with the monopole mass.

\subsection{Basic construction of a monopole}

The electromagnetic fields, solution of the wave equation $[\nabla
^{2}-\partial ^{2}/\partial (ct)^{2}]F=0$, can be described by a multipole
expansion with: 
\begin{eqnarray}
\mathbf{B}_{lm} &=&\Omega e^{-i\omega t}f_{l}(kr)\mathbf{L}Y_{lm} \\
\mathbf{E}_{lm} &=&\frac{i}{k}\mathbf{\nabla \times B}_{lm}  \nonumber
\end{eqnarray}
where $\Omega $ is a constant, $\mathbf{L}=\frac{1}{i}\mathbf{r\times \nabla 
}$, $Y_{lm}$ is the spherical harmonic of the order $l,m$ and $f_{l}$ a
Hankel plus Bessel type solution. The parameter $k$ is $\omega /c$ and $%
\left( r,t\right) $ the spherical coordinate of distance and time in the
laboratory frame. Taking the Hankel solutions, the fields $\mathbf{E}_{10}$, 
$\mathbf{B}_{10}$ can be approximated in the $kr<<1$, $r\neq 0$, region by: 
\begin{eqnarray}
\mathbf{B}_{1,0} &=&e^{-i\omega t}(-\Omega )k\mathbf{L}Y_{10}/r^{2} \\
\mathbf{E}_{1,0} &=&e^{-i\omega t}(-\Omega )\mathbf{\nabla }\left(
Y_{10}/r^{2}\right)   \nonumber
\end{eqnarray}
The $\mathbf{E}_{1,0}$ solution is exactly the field of a dipole in this
approximation and it is also written as: 
\begin{equation}
e^{-i\omega t}(-\Omega )\mathbf{\nabla }\left( Y_{10}/r^{2}\right) =(3%
\mathbf{n}(\mathbf{p}\cdot \mathbf{n})-\mathbf{p})/r^{3}
\end{equation}
with $\mathbf{p}\equiv \sqrt{3/4\pi }e^{-i\omega t}\Omega \mathbf{z}$ ($|%
\mathbf{z|=1}$) and $\mathbf{n}$ is the unitary vector directioned from the
origin to $\mathbf{r}$.

It is possible to construct a sum of solutions $\mathbf{E}_{1,0}$ (with the
corresponding magnetic field associated to it) composed over a line path
over the space. The $\mathbf{E}_{1,0}$ may also be written as $\mathbf{%
\nabla }\times [\mathbf{p\times r}/r^{3}]$ and it is possible to introduce
an elementary contribution like: 
\begin{eqnarray}
&&\mathbf{\nabla }\times [\mathbf{dp\times r}/r^{3}] \\
\mathbf{dp} &\equiv &\sqrt{3/4\pi }e^{-i\omega t}\Omega \lambda \mathbf{dl} 
\nonumber
\end{eqnarray}
where $\lambda $ is a density and $\mathbf{dl}$ the elementary oriented path
of some line in space. Defining the line path along the $z$ axis from zero
to $-L$ ($L>0,$ $L<<1/k$), the resulting electric field $\mathbf{E}_{line}$
in the region $kr<<1$ will be: 
\begin{equation}
q\left( \frac{\mathbf{r}}{4\pi r^{3}}-\frac{\mathbf{r}+L\mathbf{z}}{4\pi |%
\mathbf{r}+L\mathbf{z}|^{3}}+\delta (x)\delta (y)[\Theta (-z)-\Theta (-z-L)]%
\mathbf{z}\right) 
\end{equation}
where $\Theta (x)$ is the Heaviside function, $\delta (x)$ is the Dirac
Delta function and $q=\sqrt{3/4\pi }e^{-i\omega t}\Omega \lambda $. From
this expression it is possible to calculate the magnetic field as $\mathbf{%
\nabla \times B}=-i\omega \mathbf{E}_{line}/c$ for $kr<<1$. There are of
course other components for the fields (other terms besides the first of the
Hankel function for $x=kr<<1,$ $-(2l-1)!!(1-x^{2}/(2-4l)+...)/x^{l+1})$,
which are small as the difference $r<<1/k$ is pronounced.

Consider now an infinite set of elementary $\mathbf{E}_{line}$ solutions in
a way the lines are isotropically distributed on the radial direction with
the ends $dq$ at the origin of the coordinate system and the $-dq$ ends at
the far infinite, away from the origin. This construction defines what we
could call as a radial system of currents with spherical symmetry. In this
particular situation the first order contribution to the magnetic field is
null due to the symmetry of the construction.

Louis de Broglie proposed the similar oscillating electric charge situation,
[6], calculating the electric field in space out of the places where there
is a inward flux. Consider $A^{\mu }=(V,\mathbf{A})$ such that (in spherical
coordinates): 
\begin{eqnarray}
V &\sim &-\frac{|\mathbf{k}|^{2}}{k_{0}^{2}}\frac{e^{i(\kappa ct-|\mathbf{k}%
|r)}}{r} \\
A_{r} &\sim &\frac{i\kappa }{|\mathbf{k}|^{2}}\frac{\partial V}{\partial r} 
\nonumber \\
A_{\theta } &=&A_{\phi }=0  \nonumber
\end{eqnarray}
with $\kappa ^{2}=|\mathbf{k}|^{2}+k_{0}^{2}$, where $k_{0}$ is a parameter
related to the mass $m$ of the longitudinal field by $k_{0}=mc/\hbar $. In
guided wave problems the parameter $k_{0}$ is related to the constraint
given by the walls which guide the wave. The electric field derived from $%
A^{\mu }$ has only radial component different from zero and will be: 
\begin{eqnarray}
E_{r} &\sim &\left[ \frac{1}{r^{2}}+\frac{i|\mathbf{k}|}{r}\right]
e^{i(\kappa ct-|\mathbf{k}|r)} \\
E_{\theta } &=&E_{\phi }=0  \nonumber
\end{eqnarray}
and the magnetic field is null. If $\mathbf{k}$ approaches to zero, $\kappa $
approaches to $k_{0}$, so that in this limit: 
\begin{equation}
E_{r}\sim \frac{1}{r^{2}}e^{ik_{0}ct}
\end{equation}
which is the same result developed above for the electric field in places
outside the inward flux and with $r<<1/k$.

It interesting to make the change from the Transverse-Magnetic solutions at
the begining to the Transverse-Electric ones, $\mathbf{E}^{TM}\rightarrow -%
\mathbf{B}^{TE}$, $\mathbf{B}^{TM}\rightarrow \mathbf{E}^{TE}$, from which
we obtain the same physics for $g=-\sqrt{3/4\pi }e^{-i\omega t}\Omega
\lambda $, as we got for $q$. Now we are close to define a monopole. It is
first necessary to make $\omega \rightarrow 0$ in order the outward (inward)
magnetic flux be as constant as desidered ($g\rightarrow -\sqrt{3/4\pi }%
\Omega \lambda $). It is now necessary to create conditions for the
returning flux not to be seen. It is important to remember that the geometry
of the construction eliminates the leading contribution for the resulting
electric field. It is clear that in the given conditions the measure $|%
\mathbf{B}|^{2}-|\mathbf{E}|^{2}$ is no longer zero. In the de Broglie's
example, [6], $|\mathbf{E}|^{2}-|\mathbf{B}|^{2}\neq 0$, then the field
associated to the construction is longitudinal. If one defines the vector
potential [6], $A_{x}=A_{0}e^{i(\kappa ct-|\mathbf{k}|z)},A_{y}=A_{z}=0$,
the electric and magnetic fields result: $E_{x}=-i\kappa A_{0}e^{i(\kappa
ct-|\mathbf{k}|z)}$, $B_{y}=-i|\mathbf{k}|A_{0}e^{i(\kappa ct-|\mathbf{k}|z)}
$, and so $|\mathbf{E}|^{2}-|\mathbf{B}|^{2}=k_{0}^{2}|\mathbf{A}|^{2}$,
with $\kappa ^{2}=|\mathbf{k}|^{2}+k_{0}^{2}$, i.e., if $k_{0}\neq 0$. The
mass related to this gauge dependent object is $k_{0}\hbar /c$. In the
magnetic case, $|\mathbf{B}|^{2}-|\mathbf{E}|^{2}\neq 0$ and we arrive at
the same conclusion about the mass associated to the field since by the
present approach the charges (electric or magnetic) are defined starting
from the Maxwell equations in the vacuum, and so vector potentials can be
assigned to each field.

Now it is clear the conclusions in the pioneer work of He, Qiu and Tze, [5],
follows. In defining a monopole from the construction above, the inward flux
lines must be non-observable. The magnetic field is purelly longitudinal, so
the relation (now with $e$ as the electron's electric charge) $eg/c=n\hbar /2
$ for $g\neq 0$ ($g$ well approximated as constant in time for Physics once
we define $1/\omega \rightarrow \infty $) violates the gauge invariance
principle, a fundamental principle in Quantum Electrodynamics. If this
principle is broken by the presence of a monopole, attempts (like in
reference [3]) to make the Dirac string an artifact (non-observable) are
wrong.

\subsection{One-string description and quantization of the mass}

Now consider the line path which defines equation (8) to be defined as a
round loop - it is made when one joins the ends of the line path defined for
the elementary solution (7). In this new situation the magnetic field
results to be (for a round loop centered at the origin on the $x-y$ plane
with some radius $a$): 
\begin{eqnarray}
B_{loop}^{r} &=&\frac{Ia^{2}}{2}\frac{\cos \theta }{r^{3}} \\
B_{loop}^{\theta } &=&\frac{Ia^{2}}{4}\frac{\sin \theta }{r^{3}}  \nonumber
\end{eqnarray}

where $I\equiv -i\sqrt{3/4\pi }e^{-i\omega t}\lambda \Omega \omega /c$ and
it is considered the $a<<r<<1/k$ region.

Consider now that a solenoid made up of such loops is defined over space
(which symmetry axis follows a line over the $z$ axis) from the origin to $%
z=-\mathcal{L}$ with a density $\sigma $ of solenoids per unity distance
along its axis. Taking $\mathcal{L}<<1/k$ and $a\rightarrow 0$ the magnetic
field is: 
\begin{equation}
\mathbf{B}_{solenoid}=g\left( \frac{\mathbf{r}}{4\pi r^{3}}-\frac{\mathbf{r}+%
\mathcal{L}\mathbf{z}}{4\pi |\mathbf{r}+\mathcal{L}\mathbf{z}|^{3}}+\delta
(x)\delta (y)[\Theta (-z)-\Theta (-z-\mathcal{L})]\mathbf{z}\right)
\end{equation}

with 
\begin{equation}
g=\Upsilon \frac{\omega }{c}\sin \omega t
\end{equation}

or 
\[
g=\Upsilon \frac{\omega }{c}\cos \omega t 
\]

where $\Upsilon $ is defined to be constant and equal to $\sqrt{3/4\pi }%
a^{2}\sigma \Omega \lambda /4$ when the parameters $\sigma $ and $\lambda $
are set to compensate variations on the parameter $a$. The electric field is
null except at the solenoid (where it is infinite, i.e., not defined) in
this approximation ($r<<1/k$). 

As in the last subsection, in order to describe an oscilating magnetic
charge at the origin of the coordinate system it is necessary to define at
least a line of returning flux of magnetic field as well as to take $r<<1/k$%
. In the present construction $\mathbf{\nabla \cdot B}_{solenoid}=0$
everywhere.

Now, following the initial proposal of this section, one wants to associate
the returning magnetic flux to the gauge (arbitrary) in a way it is possible
to define a topologycal monopole. The first step towards this is to make $%
\omega \rightarrow 0$, and to make $\mathcal{L}\rightarrow \infty $. The
same problem raised in the last section happens in the one-string
description too: The measure $|\mathbf{B}|^{2}-|\mathbf{E}|^{2}$ is
different from zero and so the resulting field related to the monopole is
purelly longitudinal. If the string is not seen due to some mechanism, the
quantization condition is related to the longitudinal field only, and so,
taking $g\neq 0$ one is assuming the gauge has some physical interpretation. 

It is important to consider the case such mechanism exists. It is then
natural to consider the relation $eg/c=n\hbar /2$ ($e$ the electron's charge
- this means a gauge is chosen) when $\omega \rightarrow 0$ in order $g$
variation in time is less than our precision detection: 
\begin{equation}
g=\Upsilon \omega /c=\Upsilon mc/\hbar =n\hbar c/2e
\end{equation}

which leads to: 
\begin{equation}
m=n\hbar ^{2}/2\Upsilon e
\end{equation}

i.e., the monopole mass must be quantized if $\Upsilon $ is constant and $%
m\rightarrow 0$. This does not mean the resulting monopole mass to be
necessarily vanishing as we will see with the help of the next section. For the sake of generality, we will just quote there are approaches pointing towards a vanishing monopole mass anyway, [7]. As it was shown magnetic monopoles as well as electric
charges can be described by the present procedure, this quantization is
defined for both types of particle's charges (with $g$ instead of $e$ in the
last expression for the case of electric charge 's particles). There is a
good reason to believe $\Upsilon $ is a constant in Nature, it is probably
related to the other constants: Another interesting consideration about the
quantization condition is that, if the minimum frequence $\omega $ possible
is $2\pi c/R_{u}$, where $R_{u}$ is the physical observable Universe's
radius, and so:
\begin{equation}
\frac{\hbar c}{e}\sim \frac{\Upsilon }{R_{u}}
\end{equation}

i.e., there must be a relation between the constants of Nature (wavelenght
of a particle) and the age of the Universe (or the magnetic charge varies
with the Universe's time scale).

\subsection{Arbitrariness in the monopole description}

Other results follow from oscillating charges as in the constructions above
once a charge (magnetic or electric) at $\mathbf{r}^{\prime }$ can be
defined through the density $\rho $ by $[\rho +\rho _{0}(t)-\rho
_{0}(t)]\delta (\mathbf{r}-\mathbf{r}^{\prime })$, where $\rho _{0}$ is
function of the time and arbitrary in magnitude ($\delta $ is the
Dirac-delta function). It can be shown this arbitrariness means also no
topologycal mechanism is possible in case of monopoles in pure Quantum
Electrodynamics.

A null magnetic charge can be defined as the composition $\mathbf{A}=\mathbf{%
A}_{1}-\mathbf{A}_{2}$ (as in the first section): 
\begin{equation}
\mathbf{A}=[\mathbf{\phi }g_{0}(1-\cos \theta )/r\sin \theta ]+[\mathbf{\phi 
}g_{0}(1+\cos \theta )/r\sin \theta ]=\mathbf{\phi }g_{0}2/r\sin \theta 
\end{equation}

i.e., two monopoles of opposite charge at the same position (at the origin
of the coordinate system), with $\varepsilon <\theta <\pi -\varepsilon $ ($%
\varepsilon $ is some arbitrary vanishing angle). The magnetic flux over a
spherical surface centered at the origin is calculated as $\oint \mathbf{A}%
\cdot \mathbf{dl}$ over a closed curve around the singular string for some $%
\varepsilon <\theta <\pi /2$ ($\theta \rightarrow \varepsilon $), plus $%
-\oint \mathbf{A}\cdot \mathbf{dl}$ for a closed curve around the singular
string for $\pi -\theta $, that give us zero for the net flux. This is quite
obvious to see since the surface is open by inside and outside (the first
closed curve must be integrated on the clockwise direction while the second
in the anticlockwise direction - or vice-versa) while $\mathbf{A}$ remains
the same ($\sin \theta =\sin \pi -\theta $).

The vector potential representing a non-null magnetic charge $g$ can be
defined as the sum: 
\begin{equation}
\lbrack \mathbf{\phi }g_{0}(1-\cos \theta )/r\sin \theta ]+[\mathbf{\phi }%
g_{0}(1+\cos \theta )/r\sin \theta ]+[\mathbf{\phi }g(1-\cos \theta )/r\sin
\theta ]
\end{equation}

with $\varepsilon <\theta <\pi -\varepsilon $ ($\varepsilon $ is some
arbitrary vanishing angle). The net magnetic charge is $g$.

Now consider in the construction above $g$ is considered independent of time
but $g_{0}=g_{0}(t)$. The vector potential will be: 
\begin{equation}
\mathbf{A}=\mathbf{\phi [}g_{0}(t)2+g(1-\cos \theta )]/r\sin \theta ]
\end{equation}

and so the measure $\oint \mathbf{A}\cdot \mathbf{dl}$ around each string
will no longer be constant. The variation with time is arbitrary and so this
measure for each round loop around each string of the problem. In the case
one desires to move some of the strings off the way of particle's path, the
quantization condition cannot be applied for this particular string, and so,
it will become observable according to the picture in reference [3]. There
is an arbitrary gauge transformation instead of the one related to the Dirac
Quantization Condition. Once this condition cannot be associated to each
string but only to the net magnetic charge, conclusions in reference [3]
about the non observability of the Dirac string do not follows (the string
is not a gauge artifact which can be treated topologycally). The
arbitrariness is related to the longitudinal fields as first pointed by He,
Qiu and Tze, [5].

\section{Conclusions}

The fundamental problem now arises: Suppose there is a Dirac Monopole in
Nature. It will break gauge invariance and the Dirac string will be
observable. In what conditions has the gauge a physical interpretation and
at the same time gives a mechanism to hide the string? If the string of
sigularities is not to be seen, some physical reason (strange to pure
Quantum Electrodynamics) must be given for this to happen once now the
mechanism of Dirac or Wu-Yang, [3], fails. Is this physical reason
accounting for the gauge interpretation?

The answer relies on the longitudinal field interpretation we arrived at the
end of our attempt to built a monopole. To the longitudinal field there is
associated a mass $\omega \hbar /c^{2}$. The gauge has the physical
interpretation to be related to the mass and the charge of the monopole. Can
the mass produce the mechanism to make the returning flux non-observable? In
a recent work we attempt to show that in fact a mass can hide the Dirac
string, [8]. The gravitational effect of the mass makes not all the flat
three-dimensional world avaliable due to the distortion of the spacetime. In
the construction of fields of the last section the Hankel type of solutions
make the inward flux lines regions not defined for fields (in the
construction of fields of the one-string section the Hankel type of
solutions make the solenoid region not defined for fields and the string has
naturally some internal volume). In the reffered work, [8], we propose the
string, which is a place where the fields are not defined in the
three-dimensional flat world, is the same place the spacetime is also not
defined due to the gravitation provoked by the monopole mass.

In another recent work, [9], we propose there are Dirac monopoles in Nature
(in fact everywhere) i.e., in the electrons. It is discussed the spin
one-half electron can be defined as composed by an electric charge and two
monopoles of opposite magnetic charge performing an objetc with electric
charge and a magnetic dipole. If such construction is possible it is very
important to consider monopoles as possible entities in Nature, and so, some
interpretations about the gauge and its connection to other forces in Nature
are to be addressed.

The relation we obtain between the electric charge and the Universe's radius
is a question to be investigated in more detail, but as Feynman used to say,
once we include gravitation to the problems, it is natural to have Mach's
principle associated. The mass quantization relation is derived directly
from Dirac Quantization Condition in the case of one-string representation.
This particular case is characterized by the fact the forbidden region (the
returning flux region) has naturally some nonzero volume in space. It is
important to observe the arbitrariness in the monopole definition as defined
in section 2.3 means that the monopole mass can be non-vanishing (in fact as
big as desidered) due to the arbitrary mass of the pair $+g_{0}$, $-g_{0}$%
. In this case it is not necessary to enforce $m\rightarrow 0$ in order
to get a a large time scale for the magnetic charge oscilation (it is enough
that only $g$ varies very slowly in time, but in this case the quantization associated to the mass is lost).

It is interesting to quote the intuition Faraday had about the connection
between Gravity and Electromagnetism, [10]. He conjectured two masses in
gravitational attraction must have a solenoidal current associated to each
body. His experiments found no result: He left solenoids of various
materials in a free fall and measured the induced current due to gravity. It
is interesting the mechanism peformed to describe monopoles and then
electrons are related to gravity in our exploration: Solenoidal currents
associated to the charges in the bodies have direct relation to their
gravity.

\section{Acknowledgements}

I would like to thank Prof. C. M. G. Lattes and Prof. W. A. Rodriguez Jr 
for interesting suggestions on references and discussions. 

\section{References}
.

\par [1] P. A. M. Dirac, Proc. Roy. Soc. \textbf{A133} (1931) 60

[2] J. J. Thomson, \textit{Elements of the Mathematical Theory of
Electricity and Magnetism}, Cambridge University Press, Section 284, 3rd edition (1904).

[3] T. T. Wu and C. N. Yang, Phys. Rev. \textbf{D12} (1975) 3845

[4] R. A. Brandt and J. R. Primack, Phys. Rev. \textbf{D15} (1977) 1175

[5] H. -J. He, Z. Qiu and C. -H. Tze, Zeits. f. Phys. \textbf{C65} (1994), 
\textbf{hep-ph/9402293}

[6] L. de Broglie, \textit{Ondes \'{E}letromagn\'{e}tiqu\'{e}s \& Photons},
Gauthier-Villars Paris, Chapter 3 (1968).

[7] G. Lochak, Ann. de la Found. Louis de Broglie \textbf{9}, number 1
(1984) 5.

[8] D. S. Thober, \textit{The definition of a magnetic monopole in
Eletrodynamics combined with Gravitation}, \textbf{hep-ph/9906425}

[9] D. S. Thober, \textit{The monopoles in the structure of the electron}, 
\textbf{hep-ph/9906377}

[10] M. Faraday, \textit{Experimental Researches in Electricity}, Series
XXIV - \S\ 30, in Great Books of the Western World, volume 45. William
Benton Publisher, Encyclopaedia Britannica, Inc., Twenty-second printing, Chigago (1978).

\end{document}